\def\year{2020}\relax
\documentclass[letterpaper]{article} 
\usepackage{aaai20}  
\usepackage{times}  
\usepackage{helvet} 
\usepackage{courier}  
\usepackage[hyphens]{url}  
\usepackage{graphicx} 
\usepackage{graphicx}  
\usepackage{subfigure}
\usepackage{amsthm,amsmath,amssymb}
\usepackage{mathrsfs}
\usepackage{bm}
\usepackage{enumerate}

\title{ Multi-Feature Discrete Collaborative Filtering \\ for Fast Cold-start Recommendation}
\author{Paper ID: 7366}
\author{Yang Xu,\textsuperscript{\rm 1} Lei Zhu,\textsuperscript{\rm 1}\thanks{The corresponding author} Zhiyong Cheng,\textsuperscript{\rm 2} Jingjing Li,\textsuperscript{\rm 3} Jiande Sun\textsuperscript{\rm 1}\\
\textsuperscript{\rm 1}Shandong Normal University\\
\textsuperscript{\rm 2}Shandong Computer Science Center (National Supercomputer Center in Jinan)\\
\textsuperscript{\rm 2}Qilu University of Technology (Shandong Academy of Sciences)\\
\textsuperscript{\rm 3}University of Electronic Science and Technology of China\\
leizhu0608@gmail.com\\}
\begin{document}
\maketitle
\begin{abstract}
Hashing is an effective technique to address the large-scale recommendation problem, due to its high computation and storage efficiency on calculating the user preferences on items. However, existing hashing-based recommendation methods still suffer from two important problems: 1) Their recommendation process mainly relies on the user-item interactions and single specific content feature. When the interaction history or the content feature is unavailable (the \textit{cold-start} problem), their performance will be seriously deteriorated. 2) Existing methods learn the hash codes with relaxed optimization or adopt discrete coordinate descent to directly solve binary hash codes, which results in significant quantization loss or consumes considerable computation time. In this paper, we propose a fast cold-start recommendation method, called \textit{Multi-Feature Discrete Collaborative Filtering} (MFDCF), to solve these problems. Specifically, a low-rank self-weighted multi-feature fusion module is designed to adaptively project the multiple content features into binary yet informative hash codes by fully exploiting their complementarity. Additionally, we develop a fast discrete optimization algorithm to directly compute the binary hash codes with simple operations. Experiments on two public recommendation datasets demonstrate that MFDCF outperforms the state-of-the-arts on various aspects.
\end{abstract}

\section{Introduction}
With the development of online applications, recommender systems have been widely adopted by many online services for helping their users find desirable items. However, it is still challenging to accurately and efficiently match items to their potential users, particularly with the ever-growing scales of items and users \cite{BatmazYBK19}.

In the past, Collaborative Filtering (CF), as exemplified by Matrix Factorization (MF) algorithms \cite{DBLP:journals/computer/KorenBV09} have demonstrated great successes in both academia and industry. MF factorizes an $n\times m$ user-item rating matrix to project both users and items into a $r$-dimensional latent feature space, where the user’s preference scores for items are predicted by the inner product between their latent features. However, the time complexity for generating top-\textit{k} items recommendation for all users is $\mathcal{O}(nmr + nm\log k )$ \cite{DBLP:conf/aaai/ZhangLY17}. Therefore, MF-based methods are often computational expensive and inefficient when handling the large-scale recommendation applications \cite{ChengD0ZSK18,DBLP:journals/tois/ChengCZKK19}.

Recent studies show that the hashing-based recommendation algorithms, which encode both users and items into binary codes in Hamming space, are promising to tackle the efficiency challenge \cite{DBLP:conf/sigir/ZhangSLHLC16,DBLP:conf/sigir/ZhangWRS14}. In these methods, the preference score could be efficiently computed by Hamming distance. However, learning binary codes is generally NP-hard \cite{DBLP:journals/jacm/Hastad01} due to the discrete constraints. To tackle this problem, the researchers resort to a two-stage hash learning procedure \cite{DBLP:conf/cvpr/LiuHDL14,DBLP:conf/sigir/ZhangWRS14}: relaxed optimization and binary quantization. Continuous representations are first computed by the relaxed optimization, and subsequently the hash codes are generated by binary quantization. This learning strategy indeed simplifies the optimization challenge. However, it inevitably suffers from significant quantization loss according to \cite{DBLP:conf/sigir/ZhangSLHLC16}. Hence, several solutions are developed to directly optimizing the binary hash codes from the matrix factorization with discrete constraints. Despite much progress has been achieved, they still suffer from two problems: 1) Their recommendation process mainly relies on the user-item interactions and single specific content feature. Under such circumstances, they cannot provide meaningful recommendations for new users (e.g. for the new users who have no interaction history with the items). 2) They learn the hash codes with Discrete Coordinate Descent (DCD) that learns the hash codes bit-by-bit, which results in significant quantization loss or consumes considerable computation time.

In this paper, we propose a fast cold-start recommendation method, called \textit{Multi-Feature Discrete Collaborative Filtering} (MFDCF) to alleviate these problems. Specifically, we propose a low-rank self-weighted multi-feature fusion module to adaptively preserve the multiple content features of users into the compact yet informative hash codes by sufficiently exploiting their complementarity. Our method is inspired by the success of the multiple feature fusion in other relevant areas \cite{DBLP:journals/tcsv/WangLNYWY18,DBLP:journals/tip/WangLTLW12,DBLP:conf/mm/LuZCLNZ19,DBLP:journals/tmm/ZhuHLHSZ17}. Further, we develop an efficient discrete optimization approach to directly solve binary hash codes by simple efficient operations without quantization errors. Finally, we evaluate the proposed method on two public recommendation datasets, and demonstrate its superior performance over state-of-the-art competing baselines.

The main contributions of this paper are summarized as follows:
\begin{itemize}
\item We propose a Multi-Feature Discrete Collaborative Filtering (MFDCF) method to alleviate the cold-start recommendation problem. MFDCF directly and adaptively projects the multiple content features of users into binary hash codes by sufficiently exploiting their complementarity. To the best of our knowledge, there is still no similar work.

\item We develop an efficient discrete optimization strategy to directly learn the binary hash codes without relaxed quantization. This strategy avoids performance penalties from both the widely adopted discrete coordinate descent and the storage cost of huge interaction matrix.

\item We design a feature-adaptive hash code generation strategy to generate user hash codes that accurately capture the dynamic variations of cold-start user features. Experiments on the public recommendation datasets demonstrate the superior performance of the proposed method over the state-of-the-arts.

\end{itemize}
\section{Related Work}

In this paper, we investigate the hashing-based collaborative filtering at the presence of multiple content features for fast cold-start recommendation. Hence, in this section, we mainly review the recent advanced hashing-based recommendation and cold-start recommendation methods.

A pioneer work, \cite{DasDGR07} is proposed to exploit Locality-Sensitive Hashing (LSH) \cite{GionisIM99} to generate hash codes for Google new readers based on their item-sharing history similarity. Based on this, \cite{KaratzoglouSW10,ZhouZ12} followed the idea of Iterative Quantization \cite{GongLGP13} to project real latent representations into hash codes. To enhance discriminative capability of hash codes, de-correlation constraint \cite{DBLP:conf/cvpr/LiuHDL14} and Constant Feature Norm (CFN) constraint \cite{DBLP:conf/sigir/ZhangWRS14} are imposed when learning user/item latent representations. The above works basically follow a two-step learning strategy: relaxed optimization and binary quantization. As indicated by \cite{DBLP:conf/sigir/ZhangSLHLC16}, this two-step approach will suffer from significant quantization loss. 

To alleviate quantization loss, direct binary code learning by discrete optimization is proposed \cite{DBLP:conf/cvpr/ShenSLS15}. In the recommendation area, Discrete Collaborative Filtering (DCF) \cite{DBLP:conf/sigir/ZhangSLHLC16} is the first binarized collaborative filtering method and demonstrates superior performance over aforementioned two-stage recommendation methods. However, it is not applicable to cold-start recommendation scenarios. To address cold-start problem, on the basis of DCF, Discrete Deep Learning (DDL) \cite{DBLP:conf/wsdm/ZhangYHDYL18} applies Deep Belief Network (DBN) to extract item representation from item content information, and combines the DBN with DCF. Discrete content-aware matrix factorization methods \cite{DBLP:conf/kdd/LianLG00C17,lian2019} develop discrete optimization algorithms to learn binary codes for users and items at the presence of their respective content information. Discrete Factorization Machines (DFM) \cite{DBLP:conf/ijcai/Liu0FNLZ18} learns hash codes for any side feature and models the pair-wise interactions between feature codes. Besides, since the above binary cold-start recommendation frameworks solve the hash codes with bit-by-bit discrete optimization, they still consumes considerable computation time.

\section{The Proposed Method}
\textbf{Notations.} Throughout this paper, we utilize bold lowercase letters to represent vectors and bold uppercase letters to represent matrices. All of the vectors in this paper denote column vectors. Non-bold letters represent scalars. We denote $tr(\cdot)$ as the trace of a matrix and $\| \cdot \|_{F}$ as the Frobenius norm of a matrix. We denote $sgn(\cdot):R \to {\pm 1}$ as the round-off function.

\subsection{Low-rank Self-weighted Multi-Feature Fusion}
Given a training dataset $O={o_{i}}|^{n}_{i=1}$, which contains $n$ user's multiple features information represented with $M$ different content features (e.g. demographic information such as age, gender, occupation, and interaction preference extracted from item side information). The $m$-th content feature is ${\textbf{X}^{(m)}=[\textbf{x}^{(m)}_{1},...,\textbf{x}^{(m)}_{n}] \in \mathbb{R}^{d_{m}\times n}}$, where $d_m$ is the dimensionality of the $m$-th content feature. Since the user's multiple content features are quite diverse and heterogeneous, in this paper, we aim at adaptively mapping multiple content features $\textbf{X}^{(m)}|^{M}_{m=1}$ into a consensus multi-feature representation $\textbf{H} \in \mathbb{R}^{r \times n}$ ($r$ is the hash code length) in a shared homogeneous space. Specifically, it is important to consider the complementarity of multiple content features and the generalization ability of the fusion module. Motivated by these considerations, we introduce a self-weighted fusion strategy and then formulate the multi-feature fusion part as:
\begin{equation}
\small
\begin{split}
\mathop{\min}\limits_{\textbf{W}^{(m)},\textbf{H}} \sum \limits^{M}_{m=1}||\bm{H}-\bm{W}^{(m)}\bm{X}^{(m)}||_{F}
\end{split}
\end{equation}
\noindent where $||\cdot||_{F}$ is the Frobenius norm of the matrix. $\bm{W}^{(m)}\in \mathbb{R}^{r\times d_m},m=1,...,M$ is the mapping matrix of the $m$-th content feature, $\bm{H}\in \mathbb{R}^{r \times n}$ is the consensus multi-feature representation. According to \cite{DBLP:conf/sigir/Lu0CNZ19}, Eq.(1) is equivalent to
\begin{equation}
\small
\begin{split}
\mathop{\min}\limits_{\bm{\mu} \in \Delta_{M}, \textbf{W}^{(m)},\textbf{H}} \sum \limits^{M}_{m=1}\frac{1}{\mu^{(m)}}||\bm{H}-\bm{W}^{(m)}\bm{X}^{(m)}||^{2}_{F}
\end{split}
\end{equation}
\noindent where $\mu^{(m)}$ is the weight of the $m$-th content feature and it measures the importance of the current content feature. $\Delta_{M}\overset{def}{=}\{x\in \mathbb{R}^M | x_i \ge 0, \bm{1}^{\top}x=1\}$ is the probabilistic simplex.

In real-world recommender systems, such as Taobao\footnote{www.taobao.com} and Amazon\footnote{www.amazon.com}, there are many different kinds of users and items, which have rich and diverse characteristics. However, a specific user only has a small number of interactions in the system with limited items. Consequently, the side information of users and items would be pretty sparse. We need to handle a very high-dimensional and sparse feature matrix. To avoid spurious correlations caused by the mapping matrix, we impose a low-rank constraint on $\bm{W}$:
\begin{equation}
\small
\begin{split}
\mathop{\min}\limits_{\bm{\mu} \in \Delta_{M}, \textbf{W}^{(m)},\textbf{H}} \sum \limits^{M}_{m=1}&\frac{1}{\mu^{(m)}}||\bm{H}-\bm{W}^{(m)}\bm{X}^{(m)}||^{2}_{F}+\gamma rank(\bm{W}^{(m)})
\end{split}
\end{equation}
\noindent where $\gamma$ is a penalty parameter and $rank(\cdot)$ is the rank operator of a matrix. The low-rank constraint on $\bm{W}$ helps highlight the latent shared features across different users and handles the extremely spare observations. Meanwhile, the low-rank constraint on $\bm{W}$ makes the optimization more difficult. To tackle this problem, we adopt an explicit form of low-rank constraint as follows:
\begin{equation}
\small
\begin{split}
\mathop{\min}\limits_{\bm{\mu} \in \Delta_{M}, \textbf{W}^{(m)},\textbf{H}} \sum \limits^{M}_{m=1}&\frac{1}{\mu^{(m)}}||\bm{H}-\bm{W}^{(m)}\bm{X}^{(m)}||^{2}_{F}\\ &+\gamma \sum \limits^{M}_{m=1} \sum \limits^{l}_{i=k+1}(\sigma_{i}(\bm{W^{(m)}}))^2
\end{split}
\end{equation}
\noindent where $l$ is the total number of singular values of $\bm{W^{(m)}}$ and $\sigma_{i}(\bm{W^{(m)}})$ represents the  $i$-th singular value of $\bm{W^{(m)}}$. Note that
\begin{equation}
\small
\begin{split}
\sum \limits^{r}_{i=l+1}(\sigma_{i}(\bm{W^{(m)}}))^2 = tr({\bm{V}^{(m)}}^\top {\bm{W}^{(m)}} {\bm{W}^{(m)}}^\top {\bm{V}^{(m)}})
\end{split}
\end{equation}
\noindent where $\bm{V}$ consists of the singular vectors which correspond to the $(r-l)$-smallest singular values of ${\bm{W}^{(m)}} {\bm{W}^{(m)}}^\top$. Thus, the multiple content features fusion module can be rewritten as:
\begin{equation}
\small
\begin{split}
\mathop{\min}\limits_{\bm{\mu} \in \Delta_{M}\textbf{W}^{(m)},\textbf{H}} \sum \limits^{M}_{m=1}&\frac{1}{\mu^{(m)}}||\bm{H}-\bm{W}^{(m)}\bm{X}^{(m)}||^{2}_{F}\\ &+\gamma \sum \limits^{M}_{m=1} tr({\bm{V}^{(m)}}^\top {\bm{W}^{(m)}} {\bm{W}^{(m)}}^\top {\bm{V}^{(m)}})
\end{split}
\end{equation}
\subsection{Multi-Feature Discrete Collaborative Filtering}
In this paper, we fuse multiple content features into binary hash codes with matrix factorization, which has been proved to be accurate and scalable on addressing the collaborative filtering problems. Discrete collaborative filtering generally maps both users and items into a joint low-dimensional Hamming space where the user-item preference is measured by the Hamming similarity between the binary hash codes.

Given a user-item rating matrix $\bm{S}$ of size $n\times m$, where $n$ and $m$ are the number of users and items, respectively. Each entry $s_{ij}$ indicates the rating of a user $i$ for an item $j$. Let $b_i\in \{\pm 1\}^r$ denote the binary hash codes for the $i$-th user, and $d_j\in \{\pm 1\}^r$ denote the binary hash codes for the  $j$-th item, the rating of user $i$ for item $j$ is approximated by Hamming similarity $(\frac{1}{2}+\frac{1}{2r}b^{\top}_{i}d_j)$. Thus, the goal is to learn user binary matrix $\bm{B=[b_1,...,b_n]}\in \{\pm 1\}^{r\times n}$ and item binary matrix  $\bm{D=[d_1,...,d_m]}\in \{\pm 1\}^{r\times m}$, where $r\ll min(n,m)$ is the hash code length. Similar to the problem of conventional collaborative filtering, the basic discrete collaborative filtering can be formulated as:
\begin{equation}
\small
\begin{split}
&\mathop{\min} \limits_{\bm{B,D}}||\bm{S-B^{\top}D}||^{2}_{F}\\
&s.t. \bm{B}\in \{\pm 1\}^{r\times n}, \bm{D}\in \{\pm 1\}^{r\times m}
\end{split}
\end{equation}
To address the sparse and cold-start problem, we integrate multiple content features into the above model, by substituting the user binary feature matrix $\bm{B}$ with the rotated multi-feature representation $\bm{RH}$($\bm{R}\in \mathbb{R}^{r\times r}$ is rotation matrix) and keeping their consistency during the optimization process. The formula is given as follows:
\begin{equation}
\small
\begin{split}
&\mathop{\min}\limits_{\bm{B},\bm{D},\bm{R}}||\bm{S-H^{\top}R^{\top}D}||^{2}_{F}+\beta||\bm{B-RH}||^{2}_{F}\\
&s.t.\ \bm{B}\in\{\pm 1\}^{r\times n}, \bm{D}\in\{\pm 1\}^{r\times m}, \bm{R^{\top}R}=\bm{I_{r}}
\end{split}
\end{equation}
This formulation has three advantages: 1) Only one of the decomposed variable is imposed with discrete constraint. As shown in the optimization part, the hash codes can be learned with a simple $sgn(\cdot)$ operation instead of bit-by-bit discrete optimization used by existing discrete recommendation methods. The second regularization term can guarantee the acceptable information loss. 2) The learned hash codes can reflect user's multiple content features via $\bm{H}$ and involve the latent interactive features in $\bm{S}$ simultaneously. 3) We extract user's interactive preference from the side information of their rated items as content features. This design not only avoids the approximation of item binary matrix $\bm{D}$, reduces the complexity of the proposed model, but also effectively captures the content features of items.
\subsection{Overall Objective Formulation}
By integrating the above two parts into a unified learning framework, we derive the overall objective formulation of Multi-Feature Discrete Collaborative Filtering (MFDCF) as:
\begin{equation}
\small
\begin{split}
&\mathop{\min}\limits_{\mu^{(m)},\bm{B},\bm{D},\bm{H},\bm{R},\bm{V},\bm{W^{(m)}}} \sum\limits^{M}_{m=1} \frac{1}{\mu^{(m)}} ||\bm{H-W^{(m)}X^{(m)}}||^{2}_{F}\\
&+\alpha||\bm{S-H^{\top}R^{\top}D}||^{2}_{F}+\beta||\bm{B-RH}||^{2}_{F}\\
&+\gamma\sum\limits^{M}_{m=1} tr(\bm{{V^{(m)}}^{\top}{W^{(m)}}{W^{(m)}}^{\top}{V^{(m)}}})\\
&s.t.\ \bm{B}\in\{\pm 1\}^{r\times n}, \bm{D}\in\{\pm 1\}^{r\times m}, \bm{R^{\top}R}=\bm{I_{r}}
\end{split}
\end{equation}
\noindent where $\alpha, \beta, \gamma$ are balance parameters. The first term projects multiple content features of users into a shared homogeneous space. The second and third terms minimize the information loss during the process of integrating the multiple content features with the basic discrete CF. The last term is a low-rank constraint for $\bm{W^{(m)}}$, which can highlight the latent shared features across different users.
\subsection{Fast Discrete Optimization}
Solving hash codes in Eq.(9) is essentially an NP-hard problem due to the discrete constraint on binary feature matrix. Existing discrete recommendation methods always learn the hash codes bit-by-bit with DCD \cite{DBLP:conf/cvpr/ShenSLS15}. Although this strategy alleviates the quantization loss problem caused by conventional two-step relaxing-rounding optimization strategy, it is still time-consuming.

In this paper, with the favorable support of objective formulation, we propose to directly learn the discrete hash codes with fast optimization. Specifically, different from existing discrete recommendation methods \cite{DBLP:conf/sigir/ZhangSLHLC16,DBLP:conf/wsdm/ZhangYHDYL18,DBLP:conf/kdd/LianLG00C17,DBLP:conf/ijcai/Liu0FNLZ18}, we avoid explicitly computing the user-item rating matrix $\bm{S}$, and achieve linear computation and storage efficiency. We propose an effective optimization algorithm based on augmented Lagrangian multiplier (ALM) \cite{DBLP:journals/corr/LinCM10,DBLP:journals/technometrics/Murty07}. In particular, we introduce an auxiliary variable $\bm{Z_R}$ to separate the constraint on $\bm{R}$, and transform the objective function Eq.(9) to an equivalent one that can be tackled more easily. Then the Eq.(9) is transformed as:
\begin{equation*}
\small
\begin{split}
\mathop{\min}\limits_{\bm{\Theta}} \ \ &\sum\limits^{M}_{m=1} \frac{1}{\mu^{(m)}} ||\bm{H-W^{(m)}X^{(m)}}||^{2}_{F}\\
&+\alpha||\bm{S-H^{\top}R^{\top}D}||^{2}_{F}+\beta||\bm{B-RH}||^{2}_{F}\\
&+\gamma\sum\limits^{M}_{m=1} tr(\bm{{V^{(m)}}^{\top}{W^{(m)}}{W^{(m)}}^{\top}{V^{(m)}}})\\
&+\frac{\lambda}{2}||\bm{R-Z_{R}+\frac{G_{R}}{\lambda}}||^{2}_{F}\\
\end{split}
\end{equation*}
\begin{equation}
\small
s.t.\ \ \bm{B}\in{\pm 1}^{r\times n}, \bm{D}\in{\pm 1}^{r\times m}, \bm{R^{\top}R}=\bm{I_{r}},\bm{Z^{\top}_{R}Z_{R}}=\bm{I_{r}}
\end{equation}
\noindent where $\Theta$ denotes the variables that need to be solved in the objective function, $G_R\in \mathbb{R}^{r\times r}$ measures the difference between the target and auxiliary variable, $\lambda > 0$ is a balance parameter. With this transformation, we follow the alternative optimization process by updating each of $\bm{\mu^{(m)}},\bm{B},\bm{D},\bm{H},\bm{R},\bm{V},\bm{W^{(m)}},\bm{Z_{R}}$ and $\bm{G_{R}}$, given others fixed.

\textbf{Step 1: learning $\mu^{m}$}. For convenience, we denote $||\bm{H-W^{(m)}X^{(m)}}||_{F}$ as $h^{(m)}$. By fixing the other variables, we ignore the term that is irrelevant to $\mu^{(m)}$. The original problem can be rewritten as:
\begin{equation}
\small
\mathop{\min}\limits_{\mu^{(m)}\ge 0, \bm{1^{\top}\mu}=1} \sum\limits^{M}_{m=1} \frac{{h^{(m)}}^2}{\mu^{(m)}}
\end{equation}
With Cauchy-Schwarz inequality, we derive that
\begin{equation*}
\small
\sum \limits^{M}_{m=1}\frac{{h^{(m)}}^2}{\mu^{(m)}}\overset{(a)}{=}(\sum \limits^{M}_{m=1}\frac{{h^{(m)}}^2}{\mu^{(m)}})(\sum \limits^{M}_{m=1}\mu^{(m)})\overset{(b)}{\ge}(\sum \limits^{M}_{m=1}h^{(m)})^2
\end{equation*}
\noindent where (a) holds since $\bm{1^{\top}\mu}=1$ and the equality in (b) holds when $\sqrt{\mu^{(m)}}\varpropto \frac{h^{(m)}}{\sqrt{\mu^{(m)}}}$. Since $(\sum \limits^{M}_{m=1}h^{(m)})^2=const$, we can obtain the optimal $\mu^{(m)}$ in Eq.(11) by
\begin{equation}
\small
\mu^{(m)}=\frac{h^{(m)}}{\sum \nolimits^{M}_{m=1}h^{(m)}}
\end{equation}

\textbf{Step 2: learning $\bm{W}^{m}$}. Removing the terms that are irrelevant to the $\bm{W}^{m}$, the optimization formula is rewritten as
\begin{equation}
\small
\begin{split}
\mathop{\min}\limits_{\bm{W^{(m)}}} \sum\limits^{M}_{m=1} \frac{1}{\mu^{(m)}} &||\bm{H-W^{(m)}X^{(m)}}||^{2}_{F}\\
&+\gamma\sum\limits^{M}_{m=1} tr(\bm{{V^{(m)}}^{\top}{W^{(m)}}{W^{(m)}}^{\top}{V^{(m)}}})\\
\end{split}
\end{equation}
We calculate the derivative of Eq.(13) with respect to $\bm{W}$ and set it to zero,
\begin{equation*}
\small
\begin{split}
\sum\limits^{M}_{m=1} \frac{1}{\mu^{(m)}} &||\bm{H-W^{(m)}X^{(m)}}||^{2}_{F}\\
&+\gamma\sum\limits^{M}_{m=1} tr(\bm{{V^{(m)}}^{\top}{W^{(m)}}{W^{(m)}}^{\top}{V^{(m)}}})=0\\
\end{split}
\end{equation*}
\begin{equation}
\small
\begin{split}
\Rightarrow\gamma \bm{V^{(m)}{V^{(m)}}^{\top}W^{(m)}}+&\bm{\frac{1}{\mu^{(m)}}W^{(m)}X^{(m)}{X^{(m)}}^{\top}}\\
&\bm{=\frac{1}{\mu^{(m)}}H{X^{(m)}}^{\top}}\\
\end{split}
\end{equation}
\noindent By using the following substitutions,
\begin{equation}
\small
\left\{
\begin{array}{lr}
\bm{A=\gamma V^{(m)}{V^{(m)}}^{\top}}\\
\bm{B=\frac{1}{\mu^{(m)}}X^{(m)}{X^{(m)}}^{\top}}\\
\bm{C=\frac{1}{\mu^{(m)}}H{X^{(m)}}^{\top}}\\
\end{array}
\right.
\end{equation}
\noindent Eq.(14) can be rewritten as $\bm{AW+WB=C}$, which can be efficiently solved by Sylvester operation in Matlab.

\textbf{Step 3: learning $\bm{R}$}. Similarly, the optimization formula for updating $R$ can be represented as
\begin{equation}
\small
\begin{split}
\min\limits_{\bm{R^{\top}R=I_r}}\ &tr(-2\alpha \bm{R^{\top}DS^{\top}H^{\top}}+\alpha \bm{R^{\top}DD^{\top}RHH^{\top}}\\
&-2\beta \bm{R^{\top}BH^{\top}}-\lambda \bm{R}^{\top}(\bm{Z_R}-\frac{\bm{G_R}}{\lambda}))
\end{split}
\end{equation}
\noindent We introduce an auxiliary variable $\bm{Z_R}$ and substitute $\displaystyle\bm{R^{\top}DD^{\top}RHH^{\top}}$ with $\displaystyle\bm{R^{\top}DD^{\top}Z_RHH^{\top}}$, the Eq.(16) can be transformed into the following form
\begin{equation*}
\small
\begin{split}
\max\limits_{\bm{R^{\top}R=I_r}}\ tr(\bm{R^{\top}C})\ ,
\end{split}
\end{equation*}
\begin{equation}
\small
\begin{split}
&\bm{C}=2\alpha\bm{DS^{\top}H^{\top}}-\alpha \bm{DD^{\top}Z_RHH^{\top}}\\
&+2\beta \bm{BH^{\top}}+\lambda \bm{Z_R}-\bm{G_R}
\end{split}
\end{equation}
\noindent The optimal $\bm{R}$ is defined as $\bm{R}=\bm{PQ^{\top}}$, where $\bm{P}$ and $\bm{Q}$ are comprised of left-singular and right-singular vectors of $\bm{C}$ respectively \cite{DBLP:journals/tkde/ZhuSXC17}.

Note that, the user-item rating matrix $\bm{S}\in \mathbb{R}^{n\times m}$ is included in the term $\bm{DS^{\top}H^{\top}}$ when updating $\bm{R}$. In real-world retail giants, such as Taobao and Amazon, there are hundreds of millions of users and even more items. In consequence, the user-item rating matrix $\bm{S}$ would be pretty enormous and sparse. If we compute $\bm{S}$ directly, the computational complexity will be $\mathcal{O}(mn)$ and it is extremely expensive to calculate and store $\bm{S}$. In this paper, we apply the singular value decomposition to obtain the left singular and right singular vectors as well as the corresponding singular values of $\bm{S}$. We utilize a diagonal matrix $\Sigma^S_o$ to store the o-largest ($o\ll min\{m,n\}$) singular values, and employ an $n\times o$ matrix $\bm{P^S_o}$, an $o\times m$ matrix $\bm{Q^S_o}$ to store the corresponding left singular and right singular vectors respectively. We substitute $\bm{S}$ with $\bm{P^S_o\Sigma^S_oQ^S_o}$ and the computational complexity can be reduced to $\mathcal{O}(max\{m,n\})$.

Thus, the calculation of $\bm{C}$ can be transformed as
\begin{equation}
\small
\begin{split}
\bm{C}=&2\alpha\bm{D{Q^S_o}^{\top}{\Sigma^S_o}{P^S_o}^{\top}H^{\top}}-\alpha \bm{DD^{\top}Z_RHH^{\top}}\\
&+2\beta \bm{BH^{\top}}+\lambda \bm{Z_R}-\bm{G_R}
\end{split}
\end{equation}
\noindent With Eq.(18), both the computation and storage cost can be decreased with the guarantee of accuracy.

\textbf{Step 4: learning $\bm{H}$}. We calculate the derivative of objective function with respect to $\bm{H}$ and set it to zero, then we get
\begin{equation}
\small
\begin{split}
H=&(\sum\limits^{M}_{m=1} \frac{1}{\mu_{(m)}}\bm{I_{r}}+\alpha\bm{R^{\top}DD^{\top}R}+\beta\bm{I_{r}})^{-1} \\
&(\sum\limits^{M}_{m=1} \frac{1}{\mu_{(m)}}\bm{W^{(m)}X^{(m)}}+\alpha\bm{R^{\top}DS^{\top}}+\beta\bm{R^{\top}B})
\end{split}
\end{equation}
\noindent where $\bm{S}$ is substituted with $\bm{P^S_o\Sigma^S_oQ^S_o}$, and then we have
\begin{equation}
\small
\begin{split}
\alpha\bm{R^{\top}DS^{\top}} = \alpha\bm{R^{\top}D{Q^S_o}^{\top}{\Sigma^S_o}{P^S_o}^{\top}}
\end{split}
\end{equation}
\noindent The time complexity of computing $\bm{R^{\top}DS^{\top}}$ is reduced to $\mathcal{O}(max\{m,n\})$.

\textbf{Step 5: learning $\bm{B},\bm{D}$}. We calculate the derivative of objective function with respect to $\bm{B}$ and $\bm{D}$ respectively, and set them to zero, then we can obtain the closed solutions of $\bm{B,D}$ as
\begin{equation}
\small
\begin{split}
\bm{D}=sgn((\bm{H^{\top}R^{\top}})^{-1}\bm{S}), \ \ \ \bm{B}=sgn(\bm{RH})\\
\end{split}
\end{equation}
\noindent where $\bm{S}$ is also substituted with $\bm{P^S_o\Sigma^S_oQ^S_o}$, and update rule of $\bm{D}$ is transformed as
\begin{equation}
\small
\begin{split}
\bm{D}=sgn((\bm{H^{\top}R^{\top}})^{-1}\bm{P^S_o\Sigma^S_oQ^S_o})
\end{split}
\end{equation}

\textbf{Step 6: learning $\bm{V^{(m)}}$}. As described in Eq.(5),  $\bm{V}$ is stacked by the singular vectors which correspond to the $(r-l)$-smallest singular values of ${\bm{W}^{(m)}} {\bm{W}^{(m)}}^\top$. Thus we can solve the eigen-decomposition problem to get $\bm{V}$:
\begin{equation}
\small
\begin{split}
\bm{V}\  \leftarrow \ svd(\bm{W}^{(m)} {\bm{W}^{(m)}}^\top)
\end{split}
\end{equation}

\textbf{Step 7: learning $\bm{Z_R}$}. The objective function with respect to $\bm{Z_R}$ can be represented as
\begin{equation}
\small
\begin{split}
\max\limits_{\bm{Z_R^{\top}Z_R=I_r}}\ tr(\bm{Z_R^{\top}C_{zr}})
\end{split}
\end{equation}
\noindent where $\bm{C_{zr}}=-\alpha\bm{DD^{\top}RHH^{\top}}+\lambda\bm{R}+\bm{G_R}$. The optimal $\bm{Z_R}$ is defined as $\bm{Z_R}=\bm{P_{zr}Q_{zr}^{\top}}$, where $\bm{P_{zr}}$ and $\bm{Q_{zr}^{\top}}$ are comprised of left-singular and right-singular vectors of $\bm{C_{zr}}$ respectively.

\textbf{Step 8: learning $\bm{G_R}$}. By fixing other variables, the update rule of $\bm{G_R}$ is
\begin{equation}
\small
\begin{split}
\bm{G_R} = \bm{G_R} + \lambda(\bm{R}-\bm{Z_R})
\end{split}
\end{equation}
\subsection{Feature-adaptive Hash Code Generation for Cold-start Users}
In the process of online recommendation, we aim to map multiple content features of the target users into binary hash codes with the learned hash projection matrix $\{\bm{W^{(m)}}\}^{M}_{m=1}$. When cold-start users have no rating history in the training set and are only associated with initial demographic information, the fixed feature weights obtained from offline hash code learning cannot address the feature-missing problem.

In this paper, with the support of offline hash learning, we propose to generate hash codes for cold-start users with a self-weighting scheme. The objective function is formulated as
\begin{equation}
\small
\begin{split}
\min\limits_{\bm{B_u}\in \{\pm 1\}^{r\times n_u}} \sum^{M}_{m=1} ||\bm{B}_u-\bm{W}^{(m)}\bm{X}^{(m)}_{u}||_F
\end{split}
\end{equation}
\noindent where $\bm{W}^{(m)}\in \mathbb{R}^{r \times d}$ is the linear projection matrix from Eq.(9), $\bm{X}^{(m)}_{u}$ is content feature of target users, and $n_u$ is the number of target users. As proved by \cite{DBLP:conf/sigir/Lu0CNZ19}, Eq.(26) can be shown to be equivalent to
\begin{equation}
\small
\begin{split}
\min\limits_{\bm{B_u}\in \{\pm 1\}^{r\times n_u},\bm{\mu}\in \Delta_M} \sum^{M}_{m=1} \frac{1}{\mu^{(m)}_{u}} ||\bm{B}_u-\bm{W}^{(m)}\bm{X}^{(m)}_{u}||^2_F
\end{split}
\end{equation}

We employ alternating optimization to update $\mu^{(m)}_{u}$ and $\bm{B_u}$. The update rules are
\begin{equation}
\small
\begin{split}
&\mu^{(m)}_{u} = \frac{h^{(m)}_{u}}{\sum \nolimits^{M}_{m=1}h^{(m)}_u}\ , \ h^{(m)}_u=||\bm{B}_u-\bm{W}^{(m)}\bm{X}^{(m)}_{u}||_F\\
&\bm{B_u} = sgn(\sum^{M}_{m=1}\frac{1}{\mu^{(m)}_{u}}\bm{W}^{(m)}\bm{X}^{(m)}_{u})
\end{split}
\end{equation}
\begin{figure*}[]
	\centering
	\includegraphics[width=18cm,height=2.6cm]{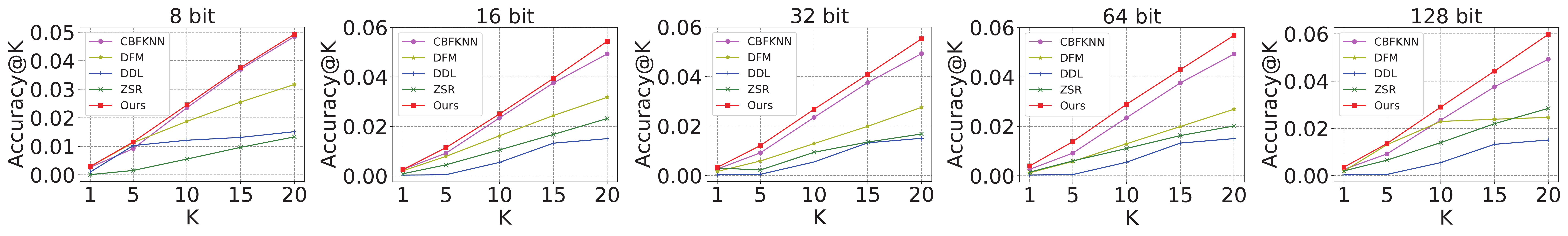}
	\vspace{-8mm}
	\caption{{Cold-start recommendation performance on MovieLens-1M.}}
	\vspace{-3mm}
	\label{Fig1}
\end{figure*}
\begin{figure*}[]
	\centering
	\includegraphics[width=18cm,height=2.6cm]{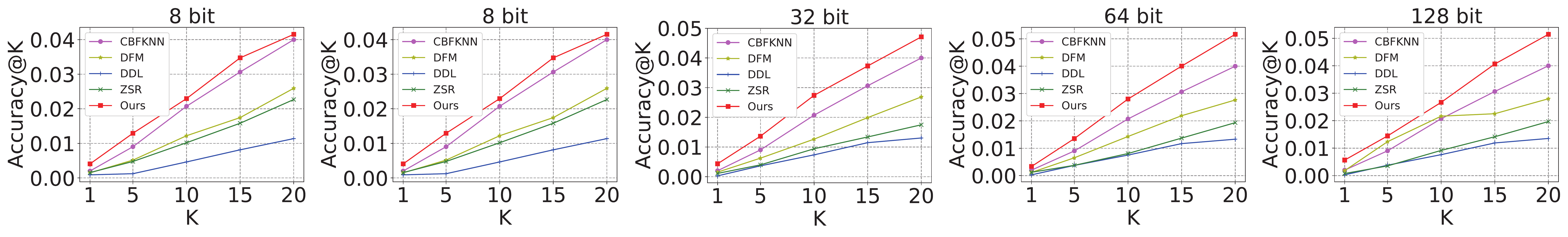}
	\vspace{-8mm}
	\caption{{Cold-start recommendation performance on BookCrossing.}}
	\vspace{-3mm}
	\label{Fig2}
\end{figure*}
\section{Experiments}
\subsection{Evaluation Datasets}
We evaluate the proposed method on two public recommendation datasets: Movielens-1M\footnote{https://grouplens.org/datasets/movielens/} and BookCrossing\footnote{https://grouplens.org/datasets/book-crossing/}. In these two datasets, each user has only one rating for an item.
\begin{itemize}
\item \textbf{Movielens-1M}: This dataset is collected from the MovieLens website by GroupLens Research. It originally includes 1,000,000 ratings from 6040 users for 3952 movies. The rating score is from 1 to 5 with 1 granularity. The users in this dataset are associated with demographic information (e.g. gender, age, and occupation), and the movies are related to 3-5 labels from a dictionary of 18 genre labels.
\item \textbf{BookCrossing}: This dataset is collected by Cai-Nicolas Ziegler from the Book-Crossing community. It contains 278,858 users providing 1,149,780 ratings (contain implicit and explicit feedback) about 271,379 books. The rating score is from 1 to 10 with 1 interval for explicit feedback, or expressed by 0 for implicit feedback. Most users in this dataset are associated with demographic information (e.g. age and location).
\end{itemize}
Considering the extreme sparsity of the original BookCrossing dataset, we remove the users with less than 20 ratings and the items rated by less than 20 users. After the filtering, there are 2,151 users, 6,830 items, and 180,595 ratings left in the BookCrossing dataset. For the MovieLens-1M dataset, we keep all users and items without any filtering. The statistics of the datasets are summarized in Table 2. The bag-of-words encoding method is used to extract the side information of the item, and one-hot encoding approach is adopted to generate feature representation of user's demographic information. To accelerate the running speed, we follow \cite{DBLP:journals/tkde/WangFHLW17} and perform PCA to reduce the interactive preference feature dimension to 128. In our experiments, we randomly select $20\%$ users as cold-start users, and their ratings are removed. We repeat the experiments with 5 random splits and report the average values as the experimental results.
\subsection{Evaluation Metrics}
The goal of our proposed method is to find out the top-$k$ items that user may be interested in. In our experiment, we adopt the evaluation metric Accuracy$@k$ \cite{DBLP:conf/wsdm/ZhangYHDYL18,Du2018Personalized} to evaluate whether the target user's favorite items appear in the top-$k$ recommendation list.

Accuracy$@k$ is to test whether the target user's favorite items that appears in the top-$k$ recommendation list.  Given the value of $k$, similar to \cite{DBLP:conf/wsdm/ZhangYHDYL18,Du2018Personalized}, we calculate Accuracy$@k$ value as:
\begin{equation}
\small
Accuracy@k \ =\ \frac{\#Hit@k}{|D_{test}|}
\end{equation}
\noindent where $|D_{test}|$ is the number of test cases, and $\#Hit@k$ is the total number of hits in the test set.
\begin{table}[]
	\centering
	\caption{Statistics of experimental datasets.}\smallskip
	\setlength{\tabcolsep}{1.5mm}{
	\begin{tabular}{ccccc}
		\hline
		Dataset & \#User & \#Item & \#Rating & Sparsity \\ \hline
		MovieLens-1M & 6,040 & 3,952 & 1,000,209 & 95.81\% \\
		BookCrossing & 2,151 & 6,830 & 180,595 & 98.77\% \\ \hline
	\end{tabular}}
\end{table}
\subsection{Evaluation Baselines}
In this paper, we compare our approach with two state-of-the-art continuous value based recommendation methods and two hashing based binary recommendation methods.
\begin{itemize}
	\item \textbf{Content Based Filtering KNN (CBFKNN)} \cite{DBLP:conf/icdm/GantnerDFRS10} is a straightforward cold-start recommendation approach based on the user similarity. Specifically, it adopts the low-dimensional projection of user attribute to calculate the user simialrity.
	\item \textbf{Discrete Factorization Machines (DFM)} \cite{DBLP:conf/ijcai/Liu0FNLZ18} is the first binarized factorization machines method that learns the hash codes for any side feature and models the pair-wise interaction between feature codes.
	\item \textbf{Discrete Deep Learning (DDL)} \cite{DBLP:conf/wsdm/ZhangYHDYL18} is a binary deep recommendation approach. It adopts Deep Belief Network to extract item representation from item side information, and combines the DBN with DCF to solve the cold-start recommendation problem.
	\item \textbf{Zero-Shot Recommendation (ZSR)} \cite{DBLP:conf/aaai/LiJL00H19} considers cold-start recommendation problem as a zero-shot learning problem \cite{DBLP:conf/cvpr/LiJLD0H19}. It extracts user preference for each item from user attribute.
\end{itemize}
In experiments, we adopt 5-fold cross validation method on random split of training data to tune the optimal hyper-parameters of all compared approaches. All the best hyper-parameters are found by grid search.
\begin{figure*}[]
	\centering
	\includegraphics[width=16cm,height=3.5cm]{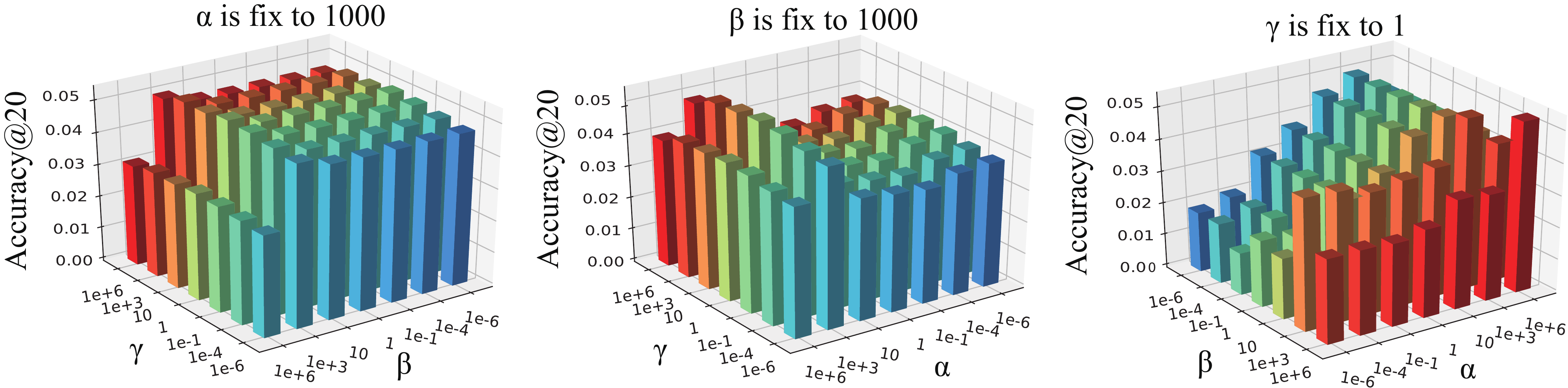}
	\vspace{-4mm}
	\caption{Performance variations with the key parameters on MovieLens-1M}
	\label{Fig3}
	\vspace{-4mm}
\end{figure*}
\begin{figure*}[t]
	\centering
	\includegraphics[width=2\columnwidth]{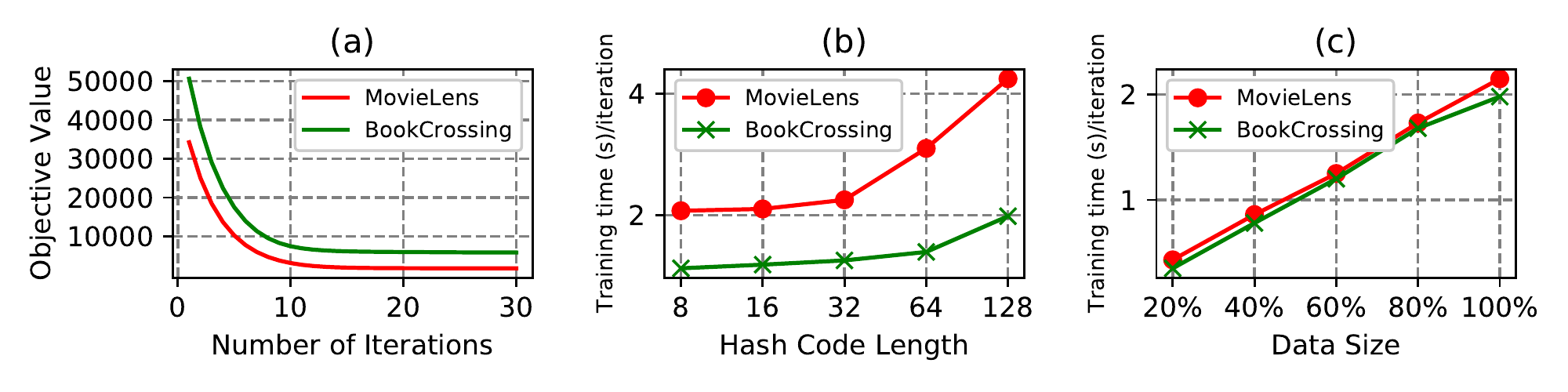}
	\vspace{-5mm}
	\caption{(a) Convergence curve, (b-c) Efficiency v.s hash code length and data size.}
	\label{fig4}
	\vspace{-5mm}
\end{figure*}
\subsection{Accuracy Comparison}
In this subseciton, we evaluate the recommendation accuracy of MFDCF and the baselines in cold-start recommendation scenario. Figure 1 and 2 demonstrate the Accuracy$@k$ of the compared approaches on two real-world recommendation datasets for the cold-start recommendation task. Compared with existing hashing-based recommendation approaches, the proposed MFDCF consistently outperforms the compared baselines. DFM exploits the factorization machine to model the potential relevance between user characteristics and product features. However, it ignores the collaborative interaction. DDL is based on the discrete collaborative filtering. It adopts DBN to generate item feature representation from their side information. Nevertheless, the structure of DBN is independent with the overall optimization process, which limits the learning capability of DDL. Additionally, these experimental results show that the proposed MFDCF outperforms the compared continuous value based hybrid recommendation methods under the same cold-start settings. The better performance of MFDCF than CBFKNN and ZSR validates the effects of the proposed multiple feature fusion strategy.
\subsection{Model Analysis}
\begin{table}[]
	\centering
	\caption{Training time comparison with two hashing-based recommendation approaches on MovieLens.}\smallskip
	\setlength{\tabcolsep}{0.7mm}{
		\begin{tabular}{cccccc}
			\hline
			Method/\#Bits & 8 & 16 & 32 & 64 & 128 \\ \hline
			DDL & 3148.86 & 3206.09 & 3289.81 & 3372.19 & 3855.81 \\
			DFM & 166.55 & 170.86 & 172.75 & 196.45 & 246.5 \\
			\textbf{Ours} & \textbf{58.87} & \textbf{60.49} & \textbf{62.55} & \textbf{68.41} & \textbf{90.17}\\ \hline
	\end{tabular}}
\end{table}
\textbf{Parameter and convergence sensitivity analysis}. We conduct experiments to observe the performance variations with the involved parameters $\alpha, \beta, \gamma$. We fix the hash code length as 128 bits and report results on MovieLens-1M. Similar results can be found on other datasets and hash code lengths. Since $\alpha, \beta$, and $\gamma$ are equipped in the same objective function, we change their values from the range of $\{10^{-6}, 10^{-4}, 10^{-1}, 1, 10, 10^{3}, 10^{6}\}$ while fixing other parameters. Detailed experimental results are presented in Figure 5. From it, we can observe that the performance is relatively better when $\alpha$ is in the range of $\{10^{6}, 10^{3}, 10\}$, $\beta$ is in the range of $\{10^{3}, 10\}$, and $\gamma$ is in the range of $\{1, 10\}$. The performance variations with $\gamma$ shows that the low-rank constraint is well on highlighting the latent shared features across different users. The convergence curves recording the objective function of MFDCF method with the number of iterations are shown in Figure 4(a). This experiment result indicates that our proposed method converges very fast.

\textbf{Efficiency v.s. hash code length and data size}. We conduct the experiments to investigate the efficiency variations of MFDCF with the increase of hash code length and training data size on two datasets. The average time cost of training iteration is shown in Figure 4(b-c). When the hash code length is fixed as 32, each round of training iteration costs several seconds and scales linearly with the increase of data size. When running MFDCF on 100\% training data, each round of iteration scales quadratically with the increase of code length due to the time complexity of optimization process is $\mathcal{O}(max\{mr^2,nr^2\})$.

\textbf{Run time comparison}. In this experiment, we compare the computation efficiency of our approach with two state-of-the-art hashing-based recommendation methods DMF and DDL. Table 2 demonstrates the training time of these methods on MovieLens-1M using a 3.4GHz Intel$\circledR$ Core(TM) i7-6700 CPU. Compared with DDL and DFM, our MFDCF is about 50 and 3 times faster respectively. The superior performance of the proposed method is attributed to that both DDL and DFM iteratively learn the hash codes bit-by-bit with discrete coordinate descent. Additionally, DDL requires to update the parameters of DBN iteratively, which consumes more time.
\section{Conclusion}
In this paper, we design a unified multi-feature discrete collaborative filtering method that projects multiple content features of users into the binary hash codes to support fast cold-start recommendation. Our model has four advantages: 1) handles the data sparsity problem with low-rank constraint. 2) enhances the discriminative capability of hash codes with multi-feature binary embedding. 3) generates feature-adaptive hash codes for varied cold-start users. 4) achieves computation and storage efficient discrete binary optimization. Experiments on two public recommendation datasets demonstrate the state-of-the-art performance of the proposed method.
\section{Acknowledgements}
The authors would like to thank the anonymous reviewers for their constructive and helpful suggestions. The work is partially supported by the National Natural Science Foundation of China (61802236, 61902223, U1836216), in part by the Natural Science Foundation of Shandong, China (No. ZR2019QF002), in part by the Youth Innovation Project of Shandong Universities, China (No. 2019KJN040), and in part by Taishan Scholar Project of Shandong, China.
\bibliography{aaai}
\bibliographystyle{aaai}
\end{document}